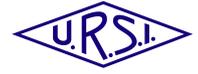

# Cassini's floating potential in Titan's ionosphere: 3-D Particle-In-Cell Simulations

Zeqi Zhang*[(1)], Ravindra Desai [(1)(2)], Oleg Shebanits [(3)], Yohei Miyake [(4)] and Hide Usui[(4)]
(1) Blackett Laboratory Imperial College London, UK; e-mail: zeqi.zhang17@imperial.ac.uk
(2) Centre for Fusion, Space & Astrophysics, University of Warwick, UK
(3) Swedish Institute of Space Physics, Uppsala, Sweden
(4) Graduate School of System Informatics, Kobe University, Kobe, Japan

## Abstract

Accurate determination of Cassini's spacecraft potential in Titan's ionosphere is important for interpreting measurements by its low energy plasma instruments. Estimates of the floating potential varied significantly, however, between the various different plasma instruments. In this study we utilize 3-D particle-in-cell simulations to understand the key features of Cassini's plasma interaction in Titan's ionosphere. The spacecraft is observed to charge to negative potentials for all scenarios considered, and close agreement is found between the current onto the simulated Langmuir Probe and that observed in Titan's ionosphere. These simulations are therefore shown to provide a viable technique for modeling spacecraft interacting with Titan's dusty ionosphere.

## 1   Introduction

Cassini's passes through the Titan's ionosphere discovered that Titan's characteristic haze was formed at much higher altitudes than previously thought [1,2]. A surprising aspect was the presence of large positively and negatively charged ions and aerosols extending up to several thousand atomic mass units (u), resulting from complex electrophyllic chemical pathways [3] and reduced photodetachment rates [4]. In order to interpret measurements by Cassini's plasma instruments of this unique environment, accurate determination of the spacecraft potential is required to correct distortions to the incoming plasma distribution function. A range of information can thus be derived, such as: composition where energy is used as a proxy for mass [1,2]; temperatures through the widths of the distributions; and ion winds from their arrival directions [5].

Cassini's spacecraft potential was however identified as different values across its various plasma instruments, with different potentials observed by Cassini's Langmuir Probe (LP) located on the end of a boom [6], and by Cassini's Plasma Spectrometer (CAPS) located on the fields and particle pallet. Different potentials were however also inferred between the co-located CAPS Ion Beam Spectrometer and Ion and Neutral Mass Spectrometer [5] and CAPS Electron Spectrometer [7].

Despite the many observational studies on Cassini's measurements in Titan's ionosphere, to our knowledge, there has not yet been a simulation study of the spacecraft interaction with its ionosphere reported in the literature, to better understand the implication for spacecraft readings as well as future spacecraft missions. In this paper, we therefore present a simulation study of the plasma interaction using 3-D Particle-In-Cell (PIC) code [8].

## 2   Simulation Technique

The Electro-Magnetic Spacecraft Environment Simulator (EMSES) [9], is a PIC simulation code developed for self-consistent three dimensional simulation of spacecraft interactions. EMSES is capable of simulating complete electromagnetic spacecraft-plasma interactions and, as a result, EMSES can capture complex plasma kinetics and variations in the vicinity of spacecraft. The electrostatic version is utilized here as the Alfvén velocities in Titan's ionosphere are significantly greater than the spacecraft velocity [10].

**Table I.** List of environmental inputs. The model set-up is otherwise identical to Zhang et al. [8].

| | |
|---|---|
| Positive ion density, $n_p$ | $528\ cm^{-3}$ |
| Positive ion mass, $m_p$ | 27 u |
| Negative ion/dust mass, $m_n$ | 47.5 u |
| Negative ion/dust density, $n_n$ | 35 cc |
| Electron density, $n_e$ | 493 cc |
| Plasma Temperature, $T_{pl}$ | 0.10 eV |
| Flow velocity, $\vec{v}_{flow}$ | (-5.64, -1.54, 2.01) $km\ s^{-1}$ |
| Magnetic field, $B_0$ | 5 nT |
| Debye length, $\lambda_D$ | 10.4 cm |

The simulation domain is 12.8 m³, across 128³ grid cells where the Cassini spacecraft is approximated using three structures: a large thin cylinder representing the antenna dish, a longer cylinder representing the main body and a small sphere representing the LP attached to the side of Cassini. This can be set at arbitrary potentials relative to the main body, and is used to collect current information.

Table I. shows the input parameters that are used for the simulation study. We analyze the night-side T29 encounter and take plasma properties estimated from Cassini's Langmuir Probe [11] at an ingress altitude of 1100 km. The dominant positive ion mass is 27 u. and, a negative ion/dust mass is taken as 200 u, and the electron depletion is taken as that inferred from the LP at 7 %.

## 3   Simulations results

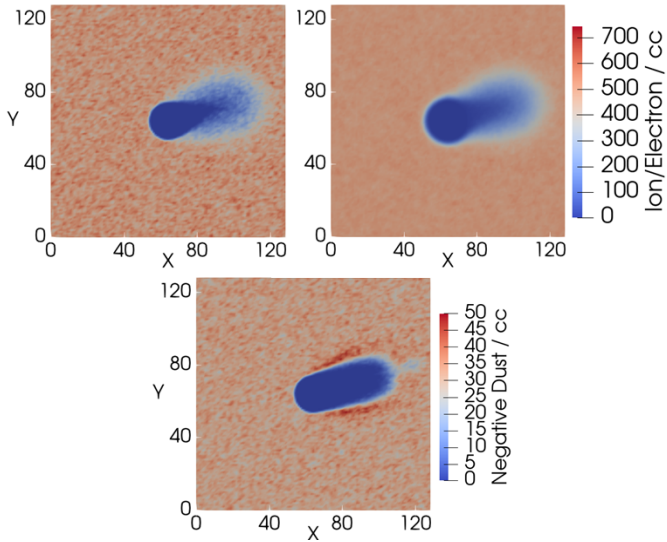

**Figure 1.** Simulation of the interaction between the Cassini spacecraft and Titan's ionosphere, showing a slice through Cassini's main body. The top-left and top-right subplots show the ion and electron number densities, respectively, and the bottom subplot the negative ion/dust number density. A full schematic of the model Cassini geometry can be found in [8].

A simulation of the interaction of Cassini spacecraft with Titan's ionospheric plasma is shown in Figure 1. No upstream disturbances are visible, such as a bow shock or electron wings; the lack of the former due to the interaction being sub-Alfvénic and the lack of the latter due to the lower magnetic field strength of the ionosphere compared to in Saturn's ionosphere where these are visible [6]. The wake from the ions is of a similar length scale to the spacecraft and the depletion tapers off rapidly behind the spacecraft, due to the relatively slow speed of the spacecraft. The electron wake is similar to the ion wake but less well defined due to their lower inertia which results in ambipolar electric fields (not shown) as they fill into the wake faster. The negative ion/dust wake on the other hand is notably longer due to their larger mass and inertia.

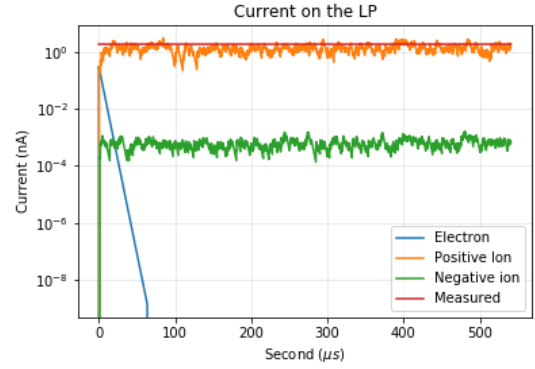

**Figure 2.** Comparison of the simulated currents onto the Langmuir Probe compared to that observed by Cassini during T29. Each simulation step corresponds to 0.033 μseconds and the probe is biased to -3 V.

The spacecraft potential is simulated at around -0.42 V, compared to the measured values of -0.62V. In the simulation, the Lagmuir probe is biased at -3V, where we would be able to collect and analyze the effect of positive ion currents on the Lagmiur probe. The comparison of the currents are shown at Figure 2. A close match between the simulated (1.5 nA) and the measured current (1.8 nA) was obtained. This is a closer match compared to comparisons to the LP currents at Saturn [6], which is attributed, in part, to a lesser presence of effective ion mass [9], heavy ions, as well as a slower spacecraft velocity.

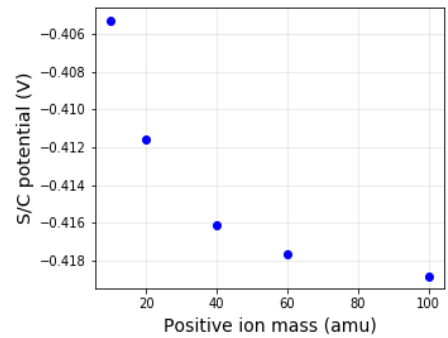

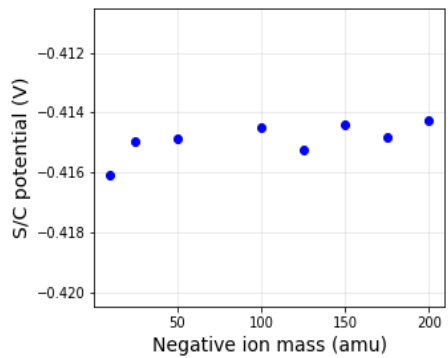

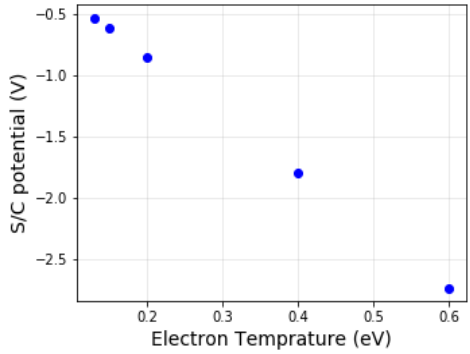

**Figure 3.** Parametric study of the effect of the positive ion mass, negative ion mass and electron temperature achieved.

To further understand the uncertainty surrounding the precise plasma parameters at Titan, we performed a parametric study on different plasma parameters to understand which are the controlling factors and how the spacecraft is subsequently affected. These results are shown at Figure 3.

First two subplots of Figure 3 demonstrate the effect of a varying positive and negative ion mass on the spacecraft potential. There produces a negligible change in the spacecraft potential when the ion masses are varied, and shows the unknown ion mass distribution function is not the controlling factor for the spacecraft potential. In contrast, in the bottom subplot, it can be seen that when the electron temperature is varied, this results in a significant deviation in the spacecraft potential. This was also observed for much larger electron depletions and indicates that the properties of the ionospheric electrons are the primary controlling factor for the spacecraft potential.

## 4 Conclusion

In this paper, we performed a simulation study of Cassini's plasma interaction within Titan's ionosphere, where large unknowns on plasma parameters remain. A good match was obtained between the simulated and observed currents onto the LP. We then varied different plasma parameters to see how the spacecraft potential is affected. When the ion mass is varied compared to the measured value, there was no significant change in the spacecraft potential, whereas when the electron temperature was varied, a significantly more negative spacecraft potential could be achieved. This change shows how accurate determination of the electron temperature in Titan's ionosphere is important to constrain and correct for spacecraft charging effects.

The simulations reported herein, highlight the main features of Cassini interacting with Titan's ionosphere but have not provided an explanation for the differing potentials observed by Cassini's various plasma instrumentation. To further analyze this phenomenon, we suggest further simulations should be conducted which consider different parts of Cassini as electrically isolated, to understand whether differential charging presents a viable explanation. Tracing the trajectory of incoming ions within this simulation may consequently provide a direct assessment of how their distribution functions have been modified by the spacecraft potential field and help to interpret Cassini measurements.

## Acknowledgements

ZZ acknowledges funding from the Royal Astronomical Society. R.T.D. acknowledges an STFC Ernest Rutherford Fellowship ST/W004801/1, NERC grants NE/P017347/1 and NE/V003062/1. YM and HU acknowledge grant no. 20K04041 from the Japan Society for the Promotion of Science: JSPS, and support from the innovative High-Performance-Computing Infrastructure (HPCI: hp210159) in Japan. OS acknowledges SNSA grant no. Dnr:195/20. This work used the Imperial College High Performance Computing Service(doi:10.14469/hpc/2232).

## References

[1] A. J. Coates, et al. "Negative ions in the Enceladus plume." *Icarus*, **206**, 2, 2010, pp. 618-622, doi:10.1016/j.icarus.2009.07.013.

[2] J. H. Waite., et al. "Chemical interactions between Saturn's atmosphere and its rings." *Science*, **362**, 6410, 2018, eaat2382, doi: 10.1126/science.aat2382

[3] T. Mihailescu et al., "Spatial Variations of Low-mass Negative Ions in Titan's Upper Atmosphere", *The Planetary Science Journal*. 1 50, 2020, doi:10.3847/PSJ/abb1ba

[4] R. T. Desai, et al. "Photodetachment and test-particle simulation constraints on negative ions in solar system plasmas." *The Planetary Science Journal*, **2**, 3, 2021, pp. 99, doi: 10.3847/PSJ/abf638

[5] F. J. Crary, et al. "Heavy ions, temperatures and winds in Titan's ionosphere: Combined Cassini CAPS and INMS observations." *Planetary and Space Science*, **57**, 14-15, 2009, pp. 1847-1856, doi: 10.1016/j.pss.2009.09.006

[6] J-E. Wahlund, et al. "Cassini measurements of cold plasma in the ionosphere of Titan." *Science*, **308**, 5724, 2005, pp. 986-989, doi: 10.1126/science.1109807

[7] R. T. Desai, et al. "Carbon chain anions and the growth of complex organic molecules in Titan's ionosphere." *Astrophysical Journal Letters*, **844**, L18, 2017, doi: 10.3847/2041-8213/aa7851

[8] Z. Zhang, et al., "Particle-in-cell simulations of the Cassini spacecraft's interaction with Saturn's

ionosphere during the Grand Finale." *Monthly Notices of the Royal Astronomical Society*, **504**, 1, 2021, pp. 964-973, doi: 10.1093/mnras/stab750

[9] Y. Miyake and H. Usui., "New electromagnetic particle simulation code for the analysis of spacecraft-plasma interactions." *Physics of Plasmas*, **16**, 6, 2009, 062904, doi: 10.1063/1.3147922

[10] S. Rehman, and R. Marchand. "Plasma-satellite interaction driven magnetic field perturbations." *Physics of Plasmas*, **21**, 9, 2014, 090701, doi: 10.1063/1.4894678

[11] O. Shebanits, et al. "Ion and aerosol precursor densities in Titan's ionosphere: A multi‑instrument case study." *Journal of Geophysical Research: Space Physics*, **121**, 10, 2016, 10-075, doi: 10.1002/2016JA022980